\documentclass[prl,twocolumn,showpacs,amsmath,amssymb]{revtex4}

\begin{document}

\title {Gravitational constant calculation methodologies}

%\date{\today}

\author {V. M. Shakhparonov}
\email{shahp@phys.msu.ru}
\affiliation {Oscillation Department, Physical Faculty, Moscow State University, Moscow 119991, Russia} 
\author{O. V. Karagioz} \author{ V. P. Izmailov}\affiliation {National Institute of Aviation Technology, 5--12 Pyrieva st., Moscow 119285, Russia}

\begin{abstract}
   We consider  the  gravitational constant calculation methodologies for a
   rectangular block of the torsion balance body presented  in  the  papers
   Phys.  Rev.  Lett.  {\bf 102},  240801 (2009) and Phys.Rev. D. {\bf 82},
   022001 (2010).  We have established the influence of non-equilibrium gas
   flows on the obtained values of $G$.
   \end{abstract}


%%%%%%%%%%%%%%%%%%%%%%%
\maketitle

  The methodologies of calculating $G$ are described in [1]. In methodology
  1, the periods of anharmonic oscillations are determined by the
  Runge-Kutta method. In methodology 2, calculations are carried out by
  analytic formulas after expanding the attraction torques in series in odd
  powers of the balance deflection angles $\varphi$. Complicated shapes of
  the bodies prevent obtaining an analytic representation of the attraction
  torques. In [2-4], the attracting bodies have a spherical shape while the
  working body is fabricated as a rectangular quartz block coated with two
  thin metal layers. In the model system, the block is changed for a thin
  rod of the same mass. Increasing the distance between the rotation axis
  and the attracting bodies provided an adequacy of the model and real
  systems at small $\varphi$. A three-position scheme of measurements has
  been considered. In position 1, the attracting bodies are placed at the
  ends of the block, in position 2 they are absent, and in position 3 they
  are rotated by $\pi/2$. The positions, oscillation periods and values of
  $G_{ij}$ according to methodologies 2 [1] and 3 [2-4] for one inverse
  measurement run (positions 3, 2, 1) and one direct run (positions 1, 2,
  3) in the first experiment as well as one direct and one inverse run in
  the second experiment are given in the table. The results are actually
  indistinguishable. Methodology 3 understates the values of $G$ in
  combinations 1-3 and 3-1 by 6 ppm, while in combinations 2-3 and 3-2 it
  slightly overstates them by a small nonlinearity persistence in the
  equations of motion. A correction of 212 ppm for inelasticity of the
  suspension thread was not taken into account. It contradicts the
  dislocation theory of inner friction [5]. After removing a calculation
  error made in [1] and using the actual values attraction torgue data from
  [4], average value of $G$ as compared with [2] in first experiment was
  overstated by 192 ppm and in second experiment was understated by 206
  ppm. In this case the first five values ​​of $G_{ij}$ understate the average value 
  of $G$ at 1120 ppm, and overstate its last eight to 386 ppm. Only this eight have 
  to be taken into account, since they reduced the standard deviation of 24 times.
  The difference in values of $G$
  for direct and inverse runs was in the first experiment 1016 ppm, and
  the second one 472 ppm. A monotone drift of the oscillation period, the
  differences in $G_{ij}$ in direct and inverse runs, appreciable
  deflections of $G_{ij}$ from the normal values when taking into account
  position 2 indicate the existence of slowly decaying non-equilibrium gas
  flows in the chamber. An unlucky choice of the material and shape of the
  working body has strengthened their effect. It could be weakened by
  increasing the density of the block and the attracting masses. The
  equality of all combinations of $G_{ij}$ was provided by an oscillation
  period diminished by 15 ms in the first experiment and increased by
  nearly 5 ms in the second experiment. One also cannot exclude the
  influence of a magnetic interaction [1], which is hard to single out
  against a more powerful factor.

\medskip
\hspace{5mm}
{\footnotesize \begin{tabular}{|cccccc|}
\hline
%%&&&&&\\
$n_i$&$n_j$&$T_i$,&$T_j$,&$10^{11}G_{ij}$,&$10^{11}G_{ij}$,\\
&&s&s&Nm/kg$^2$&Nm/kg$^2$\\
\hline
%%&&&&&\\
3&2&535.80980&535.17246&5.4845575&5.4845579\\
2&1&535.17246&532.56028&7.0498664&7.0498083\\
3&1&535.80980&532.56028&6.6787451&6.6787032\\
1&2&532.56028&535.17048&7.0445617&7.0445037\\
2&3&535.17048&535.80557&5.4652905&5.4652908\\
1&3&532.56028&535.80557&6.6701301&6.6700882\\
\hline
1&2&532.84127&535.25129&6.5380552&6.5380042\\
2&3&535.25129&536.07102&7.0682555&7.0682579\\
1&3&532.84127&536.07102&6.6640717&6.6640326\\
3&2&536.07102&535.24705&7.1049000&7.1049024\\
2&1&535.24705&532.83246&6.5506933&6.5506422\\
3&1&536.07102&532.83246&6.6824157&6.6823764\\
%%&&&&&\\
\hline
\end{tabular} }

\end{document}